\documentclass[useAMS,usenatbib]{mn2e}

\usepackage{caption}
\usepackage{graphicx}
\usepackage{textcomp}
\usepackage{latexsym}
\usepackage{varioref}
\usepackage{xspace}
\usepackage{makeidx}
\usepackage{verbatim}
\usepackage{tabularx}
\usepackage{epstopdf}
\usepackage{amsmath}
\usepackage{array}
\usepackage{units}
\usepackage{rotating}
\usepackage{float}

\def\Kepler{\textit{Kepler}}

\title[\Kepler\ spectroscopic identifications]
{Spectroscopic identifications of blue H$\alpha$ excess sources in the \Kepler\ field-of-view}
\author[S. Scaringi \textit{et al.}]
{S. Scaringi$^{1}$\thanks{E-mail: s.scaringi@astro.ru.nl}, P.J. Groot$^{1}$, K. Verbeek$^{1}$, S. Greiss $^{2}$, C. Knigge$^{3}$, E. K\"{o}rding$^{1}$\\ 
$^{1}$Department of Astrophysics/IMAPP, Radboud University Nijmegen, P.O. Box 9010, 6500 GL Nijmegen, The Netherlands \\ 
$^{2}$Department of Physics, Astronomy and Astrophysics group, University of Warwick, CV4 7AL Coventry, UK  \\ 
$^{3}$Department of Physics and Astronomy, University of Southampton, Highfield, Southampton, SO17 1BJ, UK \\ 
}

\begin{document} 

\date{}

\pagerange{\pageref{firstpage}--\pageref{lastpage}} \pubyear{2012}

\maketitle

\label{firstpage}

\begin{abstract}
We present the first results of an ongoing spectroscopic follow-up program of blue $H\alpha$ excess sources within the \Kepler\ field-of-view, in order to identify new cataclysmic variables. \Kepler\ observations of the identified targets in this work will then provide detailed, time-resolved, studies of accretion. Candidates selected from the \Kepler-INT Survey were observed with the 4.2 meter William Herschel Telescope. Out of 38 observed candidates, we found 11 new cataclysmic variables reported here for the first time, as well as 13 new quasars. Our target selection has a success rate of $29\%$ when searching for cataclysmic variables, and we show how this can be improved by including photometry obtained with the Wide-field Infrared Survey Explorer.  

\end{abstract}


\section{Introduction}

The NASA \Kepler\ mission (\citealt{borucki,haas}) was successfully launched on 2009 March 6. The \Kepler\ instrument, a 0.95 m Schmidt camera, has been developed to detect and characterise terrestrial planets within the habitable zone of stars. The observing strategy is to continuously stare at a fixed $116$ square degree field-of-view (FOV) and obtain lightcurves for target sources with a 58.8-second cadence (short cadence, SC) or with a 30-minute cadence (long cadence, LC). Thanks to its fast photometric cadence, continuous monitoring, and high photometric accuracy ($\approx2\%$ at $m_V=19$) \Kepler\ is providing unprecedented lightcurves on various types of astrophysical sources, and holds an enormous potential for other astrophysical domains on top of planet-hunting, such as asteroseismology (\citealt{gilliand}), stellar activity (\citealt{basri}), star spot monitoring (\citealt{llama}), eclipsing and close binary systems (\citealt{prsa}; \citealt{coughlin}; \citealt{bloemen11}), gyrochronology
(\citealt{meibom}), accreting white dwarfs (\citealt{fontaine}; \citealt{still10}; \citealt{cannizzo10}; \citealt{wood11}), the study of RR Lyrae stars (\citealt{benko}; \citealt{nemec}) as well as determining radial velocity amplitudes of binary systems through Doppler boosting (\citealt{vankerkwijk}; \citealt{bloemen11}; \citealt{bloemen12}). Furthermore, \Kepler\ observations are allowing for the first time detailed studies of the broad-band properties of accreting compact objects in the optical (\citealt{mushotzky}; \citealt{scaringi_mvlyra}; \citealt{scaringi_mvlyrb}).

Due to its large FOV and data-downlink rate, \Kepler\ can only monitor $\approx150,000$ targets simultaneously. Thus, in order to obtain \Kepler\ lightcurves, one needs to know in advance where the targets fall on the \Kepler\ detector pixels. It is therefore important to properly identify and classify sources within the \Kepler\ FOV in order to fully exploit the potential provided by \Kepler. To date most targets for \Kepler\ observations have been selected through the \Kepler\ Input Catalogue (KIC, \citealt{KIC}): a broad-band photometric catalogue of the field which is only complete to $m_{V}\approx16$. Given that the \Kepler\ photometric accuracy is still $2\%$ at $m_{V}=19$, identification and classification of fainter sources is also required in order to fully exploit the potential provided by \Kepler. In this respect, a new photometric survey of the \Kepler\ field has been initiated: the \Kepler-INT Survey (KIS, \citealt{greiss}). KIS is providing broad and narrow photometric measurements ($U$, $g$, $r$, $i$ and $H\alpha$) for the whole \Kepler\ FOV down to $m_g\approx20$, and is currently $\approx50\%$ complete. Thanks to KIS, it is now possible to pre-select candidates of different source types from the available photometry, and spectroscopically follow-up sources in order to obtain firm identifications in the FOV. These identification will then provide a reference for future \Kepler\ observations. 
  
Out of the different astrophysical sources in the \Kepler\ FOV, cataclysmic variables (CVs) are particularly interesting for understanding accretion disk physics. CVs are close interacting binary systems consisting of a late-type star transferring material onto a white dwarf (WD) companion via Roche-lobe overflow, and typically have orbital periods of the order of hours. Mass transfer is mostly driven by the loss of orbital angular momentum, and the transferred mass usually forms an accretion disk surrounding the central WD. Specifically, the \Kepler\ lightcurves of CVs have allowed to vigorously test the models for accretion disk dynamics that have been emerging in the past several years (\citealt{wood11, cannizzo10}). Additionally, \Kepler\ is well suited to probe the broad-band variability in CVs over a few orders of magnitude in temporal frequencies. This kind of analysis provides insight into the phenomenological similarities observed within the optical broad-band variability in accreting white dwarfs to those observed in X-rays for accreting neutron stars/black holes (\citealt{scaringi_mvlyra,scaringi_mvlyrb}).

In this paper, the initial results of a program to identify and classify blue $H\alpha$ excess sources within the \Kepler\ FOV, are presented. The results increase the number of known CVs in the field, and provide identification for other source types. The photometric candidate selection method, using the KIS catalogue, is presented in Section \ref{sec:selection}. In Section \ref{sec:observations} the spectroscopic observations, with the William Herschel Telescope (WHT), together with the data reduction procedures, are described. The results and classification of the spectra are presented in Section \ref{sec:results}. Finally, discussion and conclusions are presented in Section \ref{sec:discussion} and \ref{sec:conclusion} respectively.

\section{Candidate selection}\label{sec:selection}

Photometric candidate selection is obtained through the KIS catalogue (\citealt{greiss}), which covers about 50\% of the Kepler FOV. First, 929,732 sources classified as stellar in all 5 bands (see \citealt{gonzalez}) are selected. From these, the subset of candidate CVs is obtained by selecting all blue excess sources which also display clear $H\alpha$ excess. This is done by selecting UV-excess sources with a similar method to that used in \cite{verbeek1} and selecting H$\alpha$ emitters with a similar method to that of \cite{witham}. Objects which pass both our selection criteria are then considered to be CV candidates. We now describe the selection algorithm to independently select UV-excess sources and $H\alpha$ excess sources.

The UV-excess sources are selected from the U-g vs. g colour-magnitude diagram in the following way:
\begin{enumerate}
\item{Divide the stellar population into g-band magnitude bins. The total number of bins is determined by $N_{bins}={1\over 3} \sqrt{N_s}$, with $N_s$ being the total number of sources}
\item{Determine the bin size by keeping the ratio $\Delta g \times n_s$ approximately constant, where $n_s$ is the total number of sources in a specific bin. This will provide a balance between using small bins to precisely follow the main stellar locus, and using as many objects as possible to suppress statistical noise, resulting in variable-bin widths.}
\item{Determine the mean colour ($U-g$) in each bin, and use this as the main stellar locus}
\item{Select all UV-excess sources which lie more than one sigma away from the main stellar locus}
\end{enumerate}

The H$\alpha$ excess sources are selected in the following way from the ($r-i$) vs. ($r-H\alpha$) colour-colour diagram:
\begin{enumerate}
\item{Iteratively fit the main stellar locus with a first order polynomial using sigma clipping.}
\item{Select all H$\alpha$ excess sources which lie more than three sigma away from the fitted locus}
\end{enumerate}

\begin{figure*}
\includegraphics[width=0.45\textwidth]{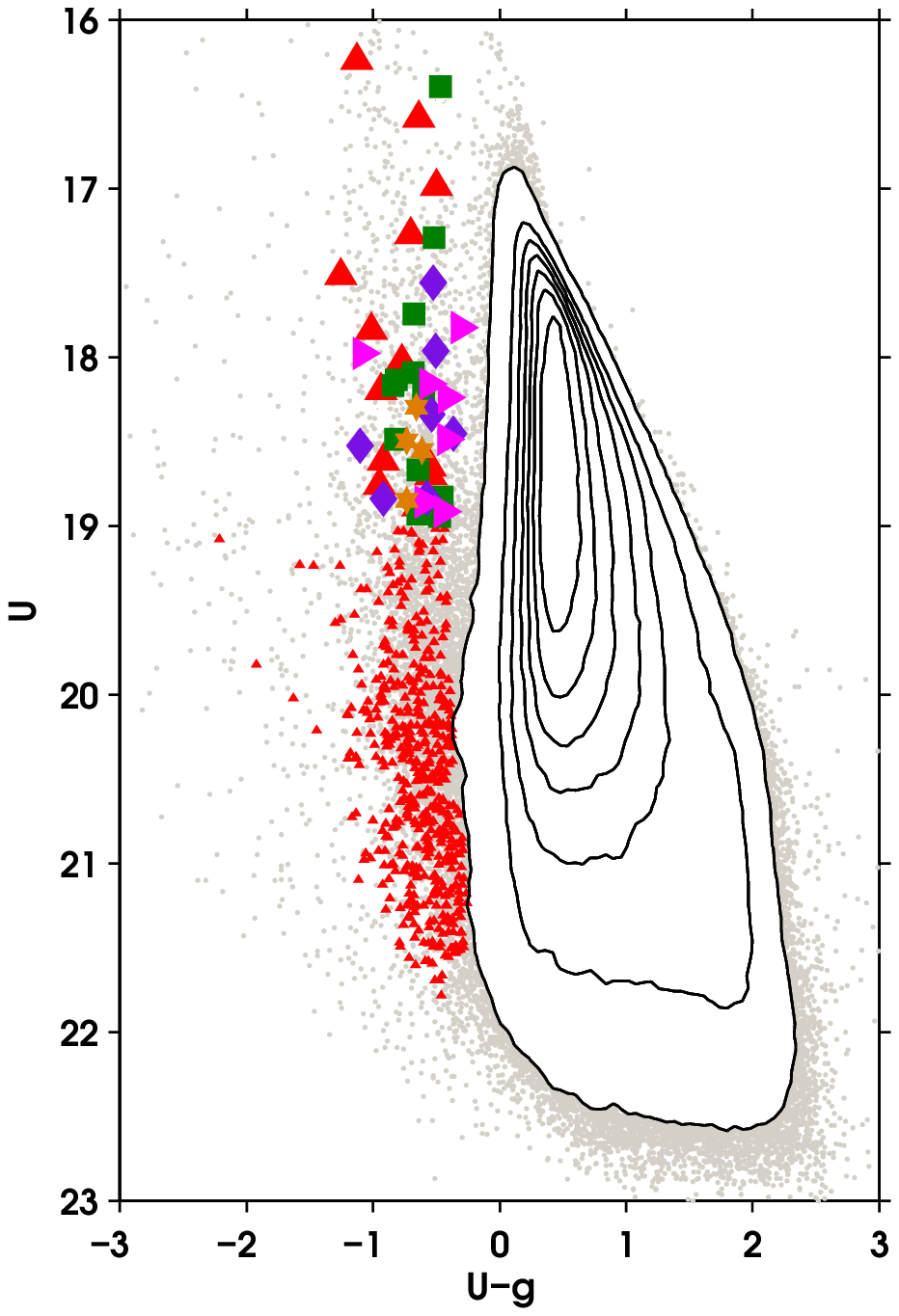}	
\includegraphics[width=0.45\textwidth]{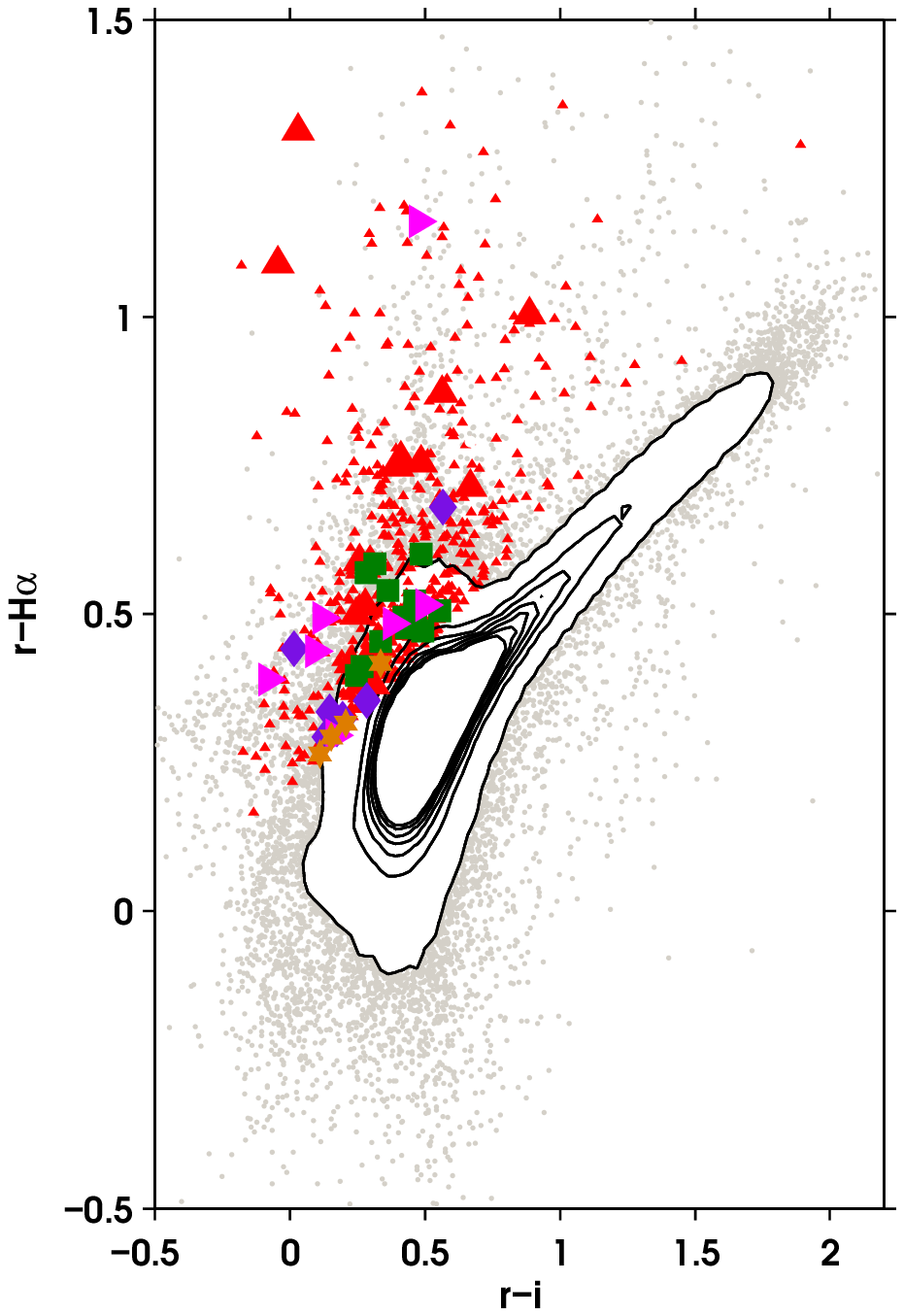}	
\caption{($U-g$) vs. $U$ and ($r-i$) vs. ($r-H\alpha$) diagram for all stellar sources in the KIS catalogue (\citealt{greiss}). Small red triangles indicate the positions of our CV candidates (see section \ref{sec:selection}). Linearly spaced isolines are shown in dense regions of the diagram. Marked with large red triangles are the positions of the identified CVs. Green squares mark the identified QSOs, magenta right-pointing triangles are stars, purple diamonds are unknowns and yellow hexagons were not observed.}
\label{fig:1}
\end{figure*}

All objects which pass both selection criteria are then considered for spectroscopic follow-up observations. In the UV-excess selection a lenient 1-sigma cut is chosen, whilst for the H$\alpha$ excess a 3-sigma cut is used. These choices are made on the basis that $H\alpha$ excess is a good indicator for CVs, but objects must still be slightly blue in colour, thus the more lenient blue selection. With this method 533 candidates ($5.7\times10^{-4}\%$ of the whole sample) are selected, shown in Fig.\ref{fig:1}. 

\section{Observations and data reduction}\label{sec:observations}

During a 7 night run, between May 30 2012 and 6 June 2012, spectra for the sample were obtained with the Intermediate dispersion Spectrograph and Imaging System (ISIS) mounted at the 4.2m William Herschel Telescope (WHT) at Roque de los Muchachos Observatory, on the island of La Palma. 

Given the high number of CV candidates, the search was restricted to the 44 candidates with $m_{U}<19$. Further a magnitude cut of $m_{U}>16$ was adopted since the KIC catalogue is complete for bright objects and also to avoid saturated photometry data. Two of the 44 candidates are already known CVs (V1504 Cyg and V363 Lyr), and spectra were obtained for 38 targets. The blue and red arms of the spectrograph were used in combination with the standard 5300 dichroic and no order sorting filter. The R300B and R316R were used in the blue and red arm, with dispersions of $0.86\AA/$pix and $0.93\AA/$pix respectively. On one night only (May 31 2012) the R600B and R600R gratings were used resulting in a dispersion of $0.45\AA/$pix and $0.49\AA/$pix respectively. The central wavelengths for the two settings are shown in Table \ref{tab:1}. The slit width (1.0-1.2 arcsec) was matched with the seeing during the observations: typically 20-30 percent larger than the seeing. All exposures were fixed at 1800 seconds, and the binning was set to 2x2 with a slow read-out speed, resulting in a signal-to-noise ratio above 20. All the WHT/ISIS spectra were reduced using IRAF\footnote{Image Reduction and Analysis Facility (IRAF) is distributed by the National Optical Astronomy Observatory, which is operated by the Association of Universities for Research in Astronomy (AURA) under cooperative agreement with the National Science Foundation.}, including bias, flat field corrections, trimming and spectral extraction. The wavelength calibration was achieved using CuNe+CuAr calibration arcs obtained every few hours during the nights. The spectra have not been corrected for telluric absorption but have been flux calibrated with standard stars taken during the same night. All reduced and calibrated spectra produced in this work can be found on-line\footnote{At http://cds.u-strasbg.fr/ and http://www.astro.ru.nl/$\sim$simo}.

\begin{table*}
\centering
\begin{tabular}{l l l l l l}
\hline
\hline
Dates       & Run & Red grating           &  Blue grating      & $\lambda_{red}$(\AA)    & $\lambda_{blue}$(\AA)  \\
\hline
30 May 2012 & 1 & R316R & R300B & 6653 & 4693 \\
31 May 2012 & 2 & R600R & R600B & 6558  & 4351 \\
2-6 June 2012 & 3 & R316R & R300B & 6656  & 4698 \\
\end{tabular}
\caption{ISIS instrument setup during the observing run.}
\label{tab:1}
\end{table*}

\section{Results}\label{sec:results}

\begin{figure*}
\includegraphics[width=0.9\textwidth, height=0.9\textheight]{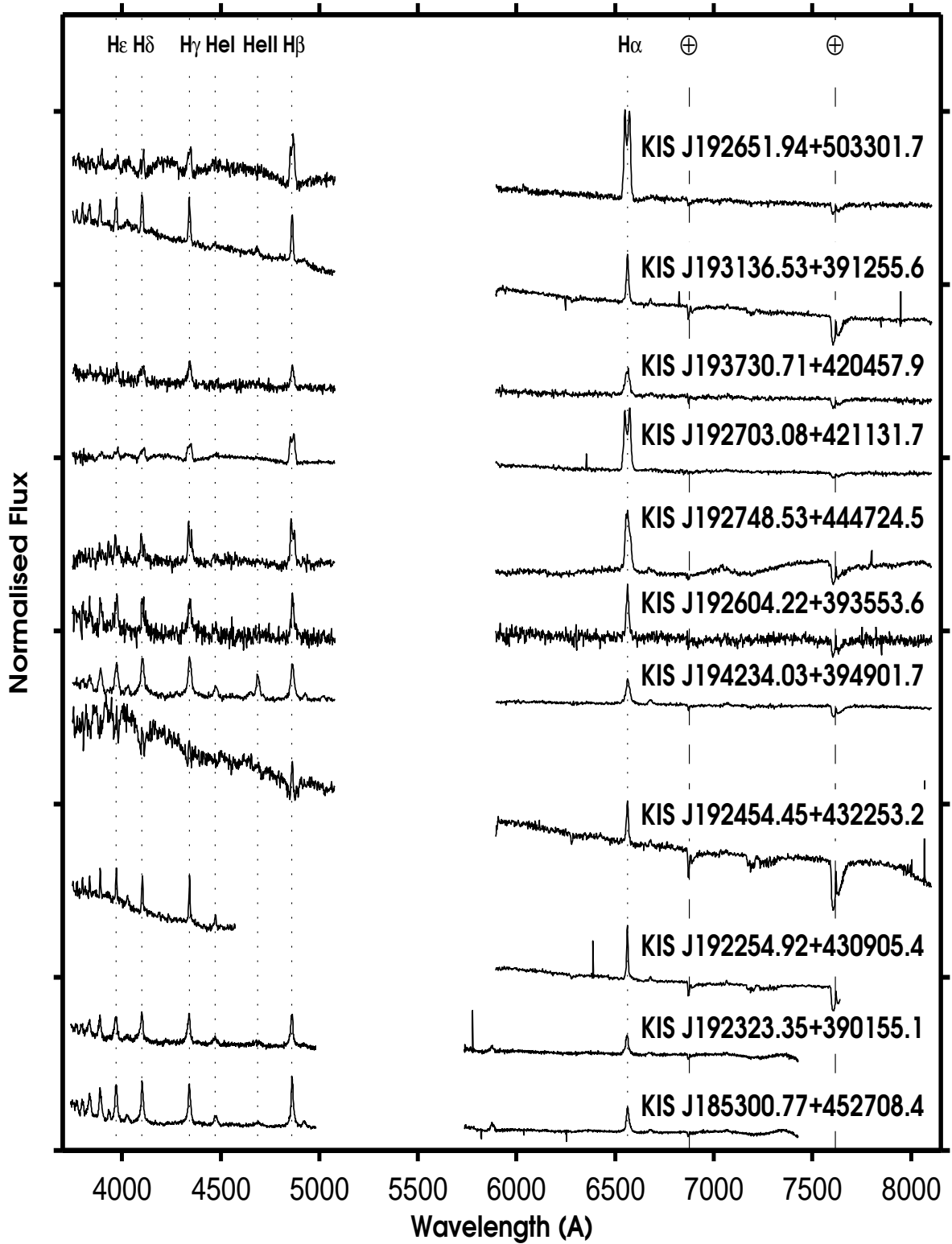}	
\caption{Flux-calibrated spectra for the 11 newly identified CVs. The spectra have all been scaled and normalised for display purposes. The position of the atmospheric A and B bands have been marked with dashed lines, whilst some Hydrogen and Helium emission lines are marked with dotted lines. The spectra can be found online.}
\label{fig:2}
\end{figure*}

The results from the spectroscopic observations are summarised in table \ref{tab:2}, as well as in Fig.\ref{fig:1} where the identified source types are coded. Out of 40 objects in our photometric sample, 11 candidates are newly identified CVs, whilst 2 were previously known. This translates to $29\%$ of the spectroscopic follow-up sample (including V1504 Cyg and V363 Lyr). The individual spectra for the CVs are shown in Fig. \ref{fig:2}. All identified CVs display clear signatures of broad Hydrogen emission similar to other known CVs. In the case of KIS J192651.94$+$503301.7 and KIS J192454.45$+$432253.2 clear evidence of underlying Balmer absorption from the accreting WD is also present, suggesting these systems have low accretion rates and possibly lie close to the orbital period minimum (\citealt{gansicke_sdss}). Additionally, KIS J194234.03$+$394901.7 is the only CV in our sample showing clear HeII 4686 \AA\ emission, suggesting that the accreting WD is strongly magnetic.

\begin{figure*}
\includegraphics[width=0.9\textwidth, height=0.9\textheight]{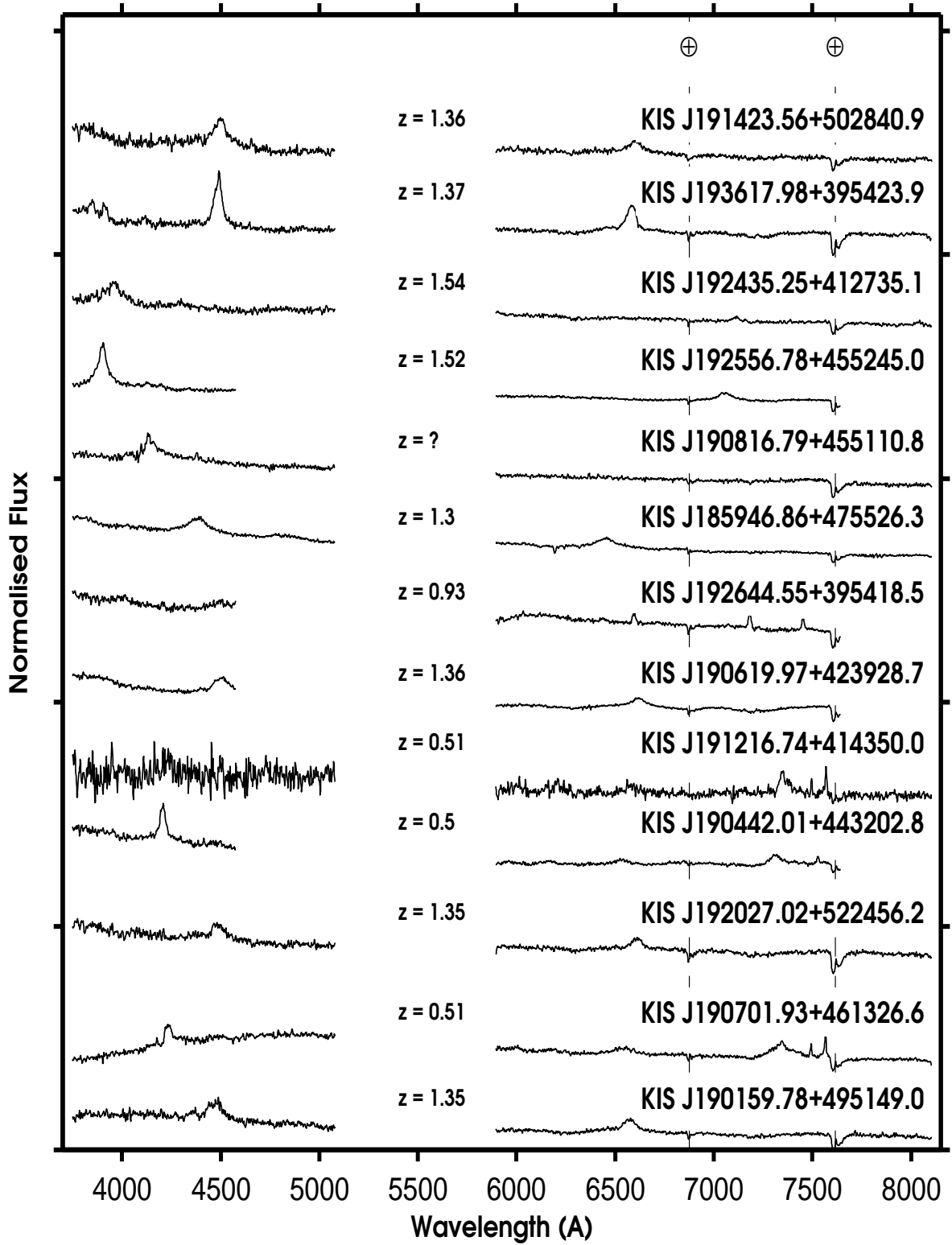}	
\caption{Flux-calibrated spectra for the 13 newly identified QSOs. The spectra have all been scaled and normalised for display purposes. The position of the atmospheric A and B bands have been marked with dashed lines and the inferred redshift for each QSO is shown. The spectra can be found online.}
\label{fig:3}
\end{figure*}

13 candidates have been firmly classified as QSOs, and the obtained redshift for these can be found in table. \ref{tab:2}, as well as in Fig. \ref{fig:3}. The inferred redshifts are based on the clear detection of at least 2 broad emission lines. In one case (KIS J190816.79$+$455110.8) we only detect one emission line, and cannot infer a definite redshift. Two possible identifications for this line are considered. The first is that the line is MgII 2800 \AA, and the QSO does not display any $H\beta$ or OIII lines, resulting in $z\approx0.5$. The second option is that the visible line is CII 1336 \AA, with MgII falling within the wavelength gap, resulting in $z\approx1$. All identified QSOs have passed the selection criteria as they are intrinsically blue, and/or one or more broad/narrow emission lines lie close to the $H\alpha$ band. Because of this the QSOs all have similar redshifts clustered around $z\approx0.5,1.3,1.5$. One exception is the QSO KIS J192644.55+395418.5 at $z=0.9$, which shows few narrow emission lines. This is the only QSO in the sample displaying NeV 3347 \AA\ and 3427 \AA\ (falling in the $H\alpha$ band in the observed frame) as well as OII 3728 \AA, and MgII 2800 \AA\ falling within the wavelength gap. A brief literature search showed this object to also display strong extended radio emission (NVSS J192644$+$395418 or 4C $+39.60$, $\approx600$mJy, \citealt{radioP}) on a few arcseconds scale. 

\begin{figure*}
\includegraphics[width=0.9\textwidth, height=0.9\textheight]{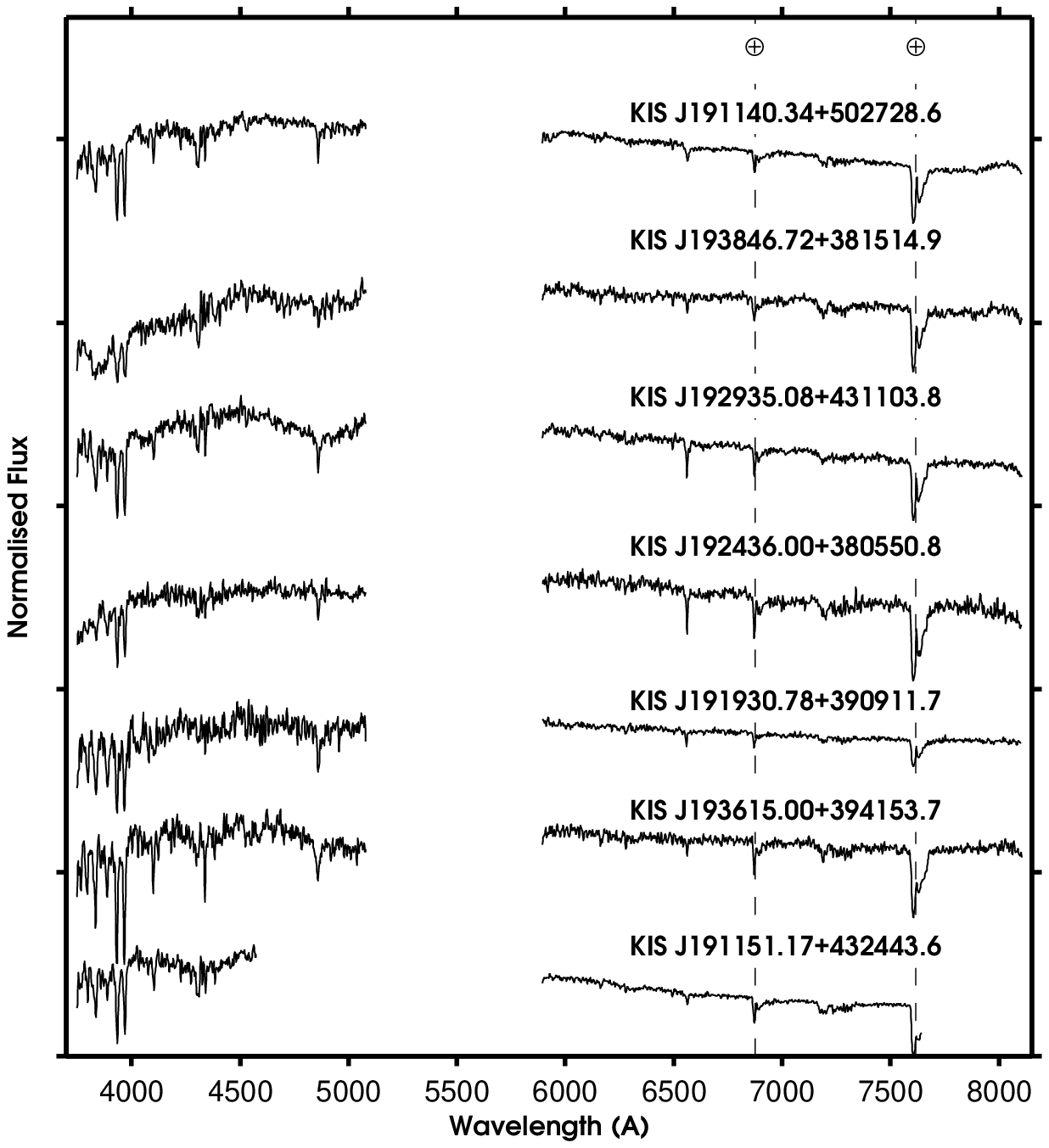}	
\caption{Flux-calibrated spectra for the 7 stars in our sample. The spectra have all been scaled and normalised for display purposes. The position of the atmospheric A and B bands have been marked with dashed lines. The spectra can be found online.}
\label{fig:4}
\end{figure*}

\begin{figure*}
\includegraphics[width=0.9\textwidth, height=0.9\textheight]{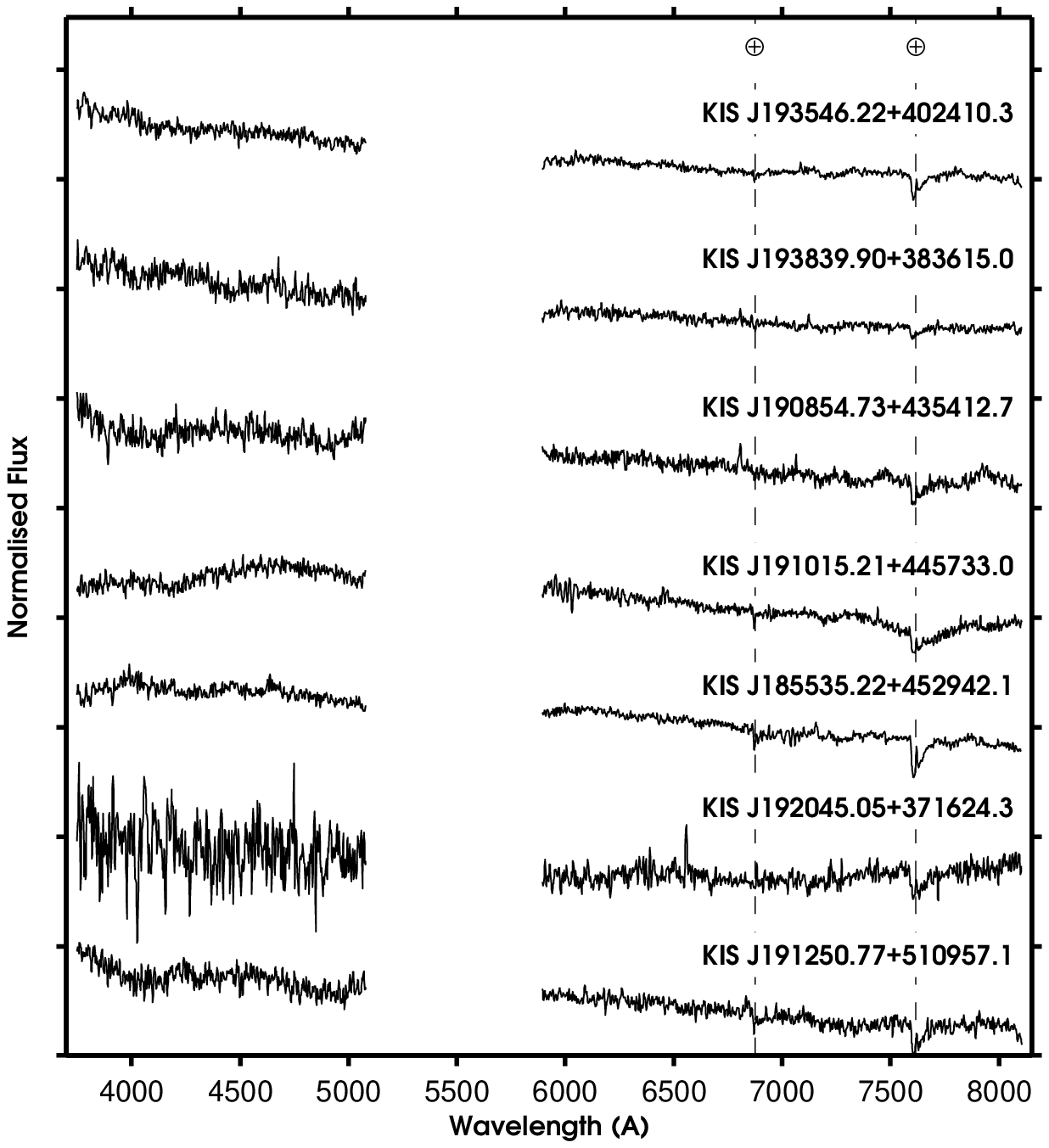}	
\caption{Flux-calibrated spectra for the 7 unidentified spectra in our sample. The spectra have all been scaled and normalised for display purposes. The position of the atmospheric A and B bands have been marked with dashed lines. The spectra can be found online.}
\label{fig:5}
\end{figure*}

All of the 7 identified stars are shown in Fig. \ref{fig:4}, and lie close to the ($U-g$) boundary and the ($r-H\alpha$) stellar locus of the photometric selection (see Fig. \ref{fig:1}), with the exception of KIS J192436.00$+$380550.8, with ($r-H\alpha$)$=1.16$. Since no $H\alpha$ emission is detected from the spectrum it is possible that this star is variable. All other identified stars have been included in the spectroscopic follow-up sample because of photometric scatter within the KIS data.

Fig. \ref{fig:5} displays the spectra of the 7 unidentified sources showing a relatively featureless continuum. One possibility is that all, or most, of these sources are also QSOs, where the broad emission lines happen to fall within the wavelength gap ($\approx5100$\AA-$5800$\AA) and outside the wavelength range. If this were the case, the redshift for this sample would be clustered around $z\approx1$, with the MgII 2800 \AA\ emission line falling in the wavelength gap. Another possibility for these sources is that they are DC WDs, which also display featureless continua. However, from the spectra in Fig. \ref{fig:5} alone, it is impossible to firmly determine the nature of these sources. Four objects in the spectroscopic follow-up sample were not observed, marked with dashes in table \ref{tab:2}.

\begin{table*}
\centering
\begin{tabular}{l l l l l l}
\hline
\hline
KIS ID   & KIC ID  & KIC magnitude & Run & Class & Comments  \\

\hline 
\hline 
J185300.77+452708.4  &   9071514  &   18.664  &   2  &   CV &    \\   
J185535.22+452942.1  &   9072731  &   18.057  &   3  &   ? &   - \\   
J185946.86+475526.3  &   10651040 &   17.525  &   1  &   QSO &   1.30 \\   
J190159.78+495149.0  &   11701965 &   18.231  &   3  &   QSO &   1.35 \\   
J190442.01+443202.8  &   8482611  &   18.618  &   3  &   QSO &   0.50 \\   
J190619.97+423928.7  &   7102641  &   18.601  &   3  &   QSO &   1.36 \\   
J190701.93+461326.6  &   9577724  &   16.684  &   3  &   QSO &   0.51 \\   
J190816.79+455110.8  &   9332324  &   19.133  &   1  &   QSO &   ? \\   
J190851.59+430031.2  &   7431243  &   16.672  &   -  &   CV &   V363 Lyr \\   
J190854.73+435412.7  &   8086137  &   18.766  &   3  &   ? &   - \\   
J191015.21+445733.0  &   8743676  &   18.094  &   1  &   ? &   - \\   
J191140.34+502728.6  &   12006762 &   17.518  &   3  &   STAR &   G \\   
J191151.17+432443.6  &   7743065  &   17.122  &   3  &   STAR &   FGK \\   
J191155.15+493850.0  &      -     &   -       &   -  &   - & -   \\   
J191216.74+414350.0  &   6353292  &   18.956  &   3  &   QSO &   0.51 \\   
J191228.24+473844.7  &   10461685 &   18.441  &   -  &   - &   - \\   
J191250.77+510957.1  &   12351437 &   17.669  &   3  &   ? &   - \\   
J191423.56+502840.9  &   12008037 &   18.966  &   3  &   QSO &   1.36 \\   
J191712.48+461936.9  &   9643342  &   18.958  &   -  &   - &   - \\   
J191852.60+424227.2  &   7193516  &   18.901  &   -  &   - &   - \\   
J191930.78+390911.7  &   4051140  &   17.801  &   3  &   STAR &   F \\   
J192027.02+522456.2  &   12984288 &   17.577  &   3  &   QSO &   1.35 \\   
J192045.05+371624.3  &   1715935  &   19.807  &   3  &   ? &   - \\   
J192254.92+430905.4  &   7524178  &   16.693  &   3  &   CV &     \\   
J192323.35+390155.1  &   3952037  &   17.338  &   2  &   CV &     \\   
J192435.25+412735.1  &   6121502  &   18.703  &   3  &   QSO &   1.54 \\   
J192436.00+380550.8  &   2850987  &   17.637  &   1  &   STAR &   FGK \\   
J192454.45+432253.2  &   7680833  &   17.214  &   3  &   CV &     \\   
J192556.78+455245.0  &   9341505  &   18.746  &   3  &   QSO &   1.52 \\   
J192604.22+393553.6  &      -     &   -       &   1  &   CV &     \\   
J192644.55+395418.5  &   4835238  &   18.785  &   3  &   QSO &   0.93 \\   
J192651.94+503301.7  &   12062071 &   18.927  &   3  &   CV &     \\   
J192703.08+421131.7  &      -     &   -       &   1  &   CV &    \\   
J192748.53+444724.5  &   8625249  &   18.353  &   3  &   CV &     \\   
J192856.46+430537.4  &   7446357  &   15.805  &   -  &   CV &   V1504 Cyg \\   
J192935.08+431103.8  &   7529625  &   18.350  &   3  &   STAR &   G \\   
J193136.53+391255.6  &   4163085  &   17.663  &   1  &   CV &    \\   
J193546.22+402410.3  &   5282446  &   19.087  &   3  &   ? &   - \\   
J193615.00+394153.7  &      -     &   -       &   3  &   STAR &   F \\   
J193617.98+395423.9  &   4843809  &   18.447  &   3  &   QSO &   1.37 \\   
J193730.71+420457.9  &   6615102  &   18.568  &   3  &   CV &    \\   
J193839.90+383615.0  &   3557541  &   19.350  &   3  &   ? &   - \\   
J193846.72+381514.9  &      -     &   -       &   3  &   STAR &   G \\

\hline

\end{tabular}
\caption{List and identifications of the photometric CV candidates with $m_{U}<19$ obtained from the selection described in Section \ref{sec:selection} sorted by right ascension. Two objects in the sample (V1504 Cyg and V363 Lyr) already had previous identifications and were not observed, as well as objects whose type is denoted by a dash. Where the object is found to be a QSO, the comments column shows the inferred redshift. Other source types show additional information on the classification. Observed but unidentified objects are marked with a question mark.}
\label{tab:2}
\end{table*}

\section{Discussion}\label{sec:discussion}

Although the spectroscopic CV recovery rate from the photometric selection method ($29\%$) is reasonable, it is hard from the diagrams in Fig. \ref{fig:1} to improve on this rate as there is no clear distinction between types (see also \citealt{verbeek12}). One way to overcome this degeneracy is to include additional colours from the Wide-Field Infrared Survey Explorer (WISE, \citealt{WISE,cutri}).

\begin{figure*}
\includegraphics[width=0.45\textwidth]{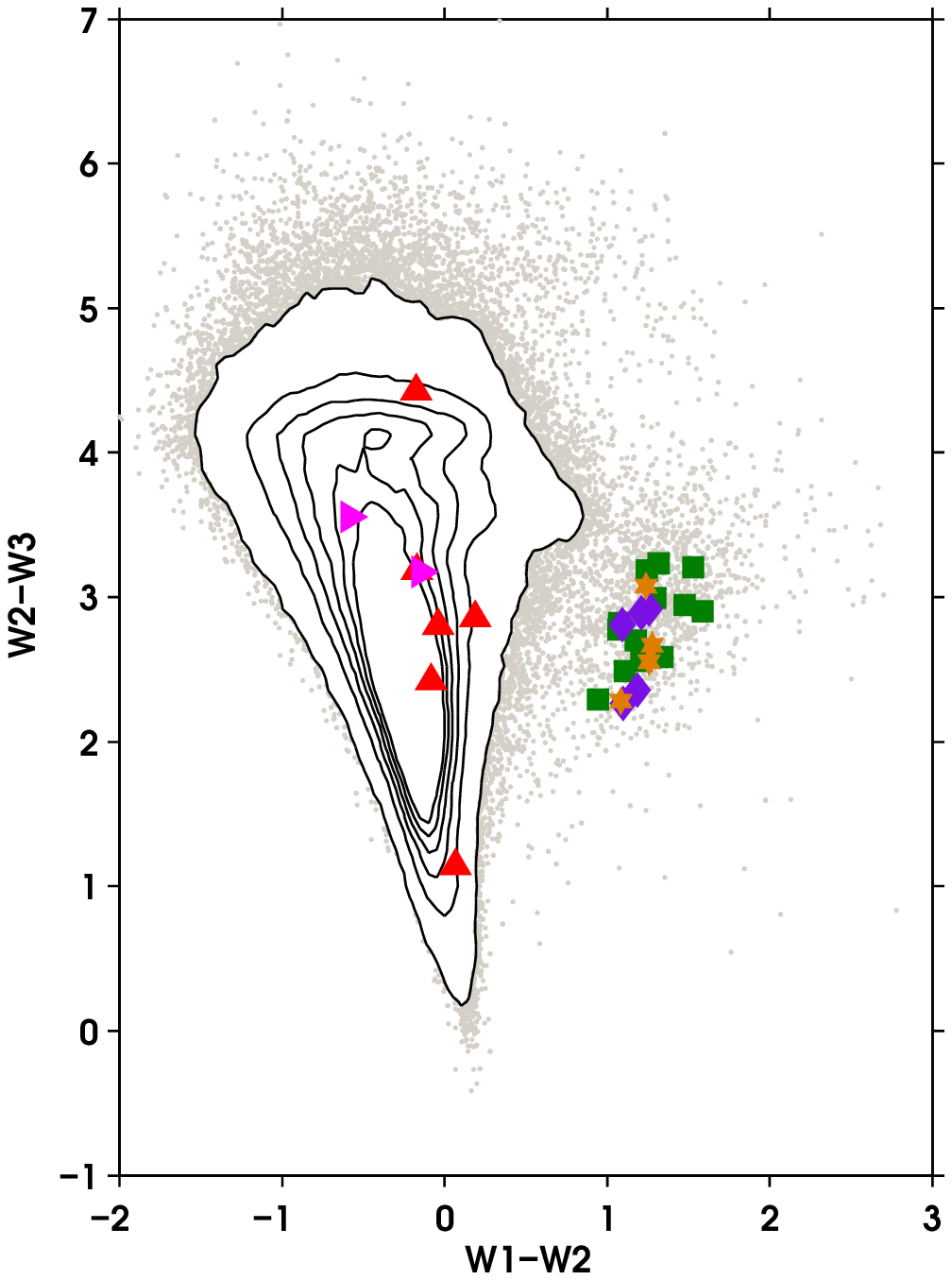}	
\includegraphics[width=0.45\textwidth]{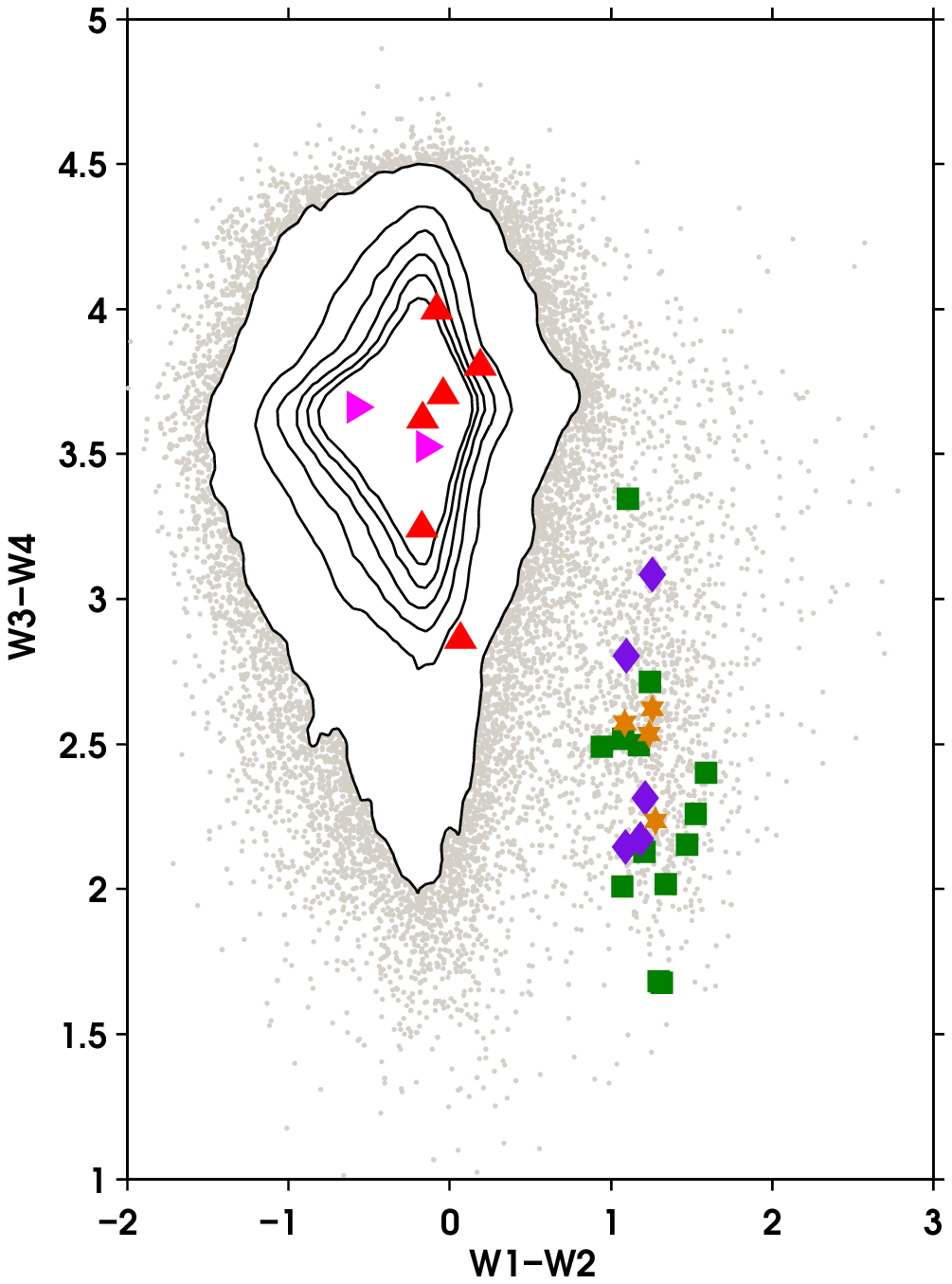}	
\caption{WISE colour-colour diagrams for stellar objects matched with the KIS catalogue. Linearly spaced isolines are shown in dense regions. Also shown are the locations of our observed sample with the same colour coding as in Fig.\ref{fig:1}. Fewer objects are present than in Fig.\ref{fig:1} due to missing WISE counterparts.}
\label{fig:6}
\end{figure*}

To show this, all stellar sources from the KIS catalogue and the WISE catalogue (\citealt{cutri}) were matched with a $1^{\prime\prime}$ radius, resulting in $345,718$ matches. Of these, 30 objects were present in our spectroscopic follow-up sample of 44. Fig. \ref{fig:6} shows the WISE colour-colour diagrams with the positions of the spectroscopic identifications. One important property of the matched 30 objects is that most of these are QSOs. In fact, matches are found for all 13 QSOs, but only 6 CVs. Furthermore, Fig. \ref{fig:6} shows how CVs and QSOs have intrinsically different WISE colours. Future spectroscopic observations to identify more CVs (or QSOs) in the \Kepler\ field should obtain a better recovery rate by using both KIS and WISE data.

5 out of the 7 unidentified sources from Fig. \ref{fig:5}, as well as the 4 unobserved objects in our photometric sample, fall close to the identified QSOs in Fig. \ref{fig:6}. Additionally, the possibility that some of these are WDs can be partially ruled out from the work of \cite{debes}, which have studied the sample of SDSS WDs in the WISE data. Specifically, the mean ($W1-W2$) WISE colours of WDs is $\approx0.15$, with very few objects owning ($W1-W2$)$>1$. All of the 9 objects without firm identifications in Fig.\ref {fig:5} have ($W1-W2$)$>1$, and are thus very probably QSOs. Two objects without WISE matches (KIS J193839.90$+$383615.0 and KIS J191250.77$+$510957.1) could still potentially be WDs.

One important measurement to understand the population and evolution of CVs, as well as other accreting compact objects, is the space density of CVs. To this end, one of the potential goals of this spectroscopic follow-up campaign is to infer the CV space density within the \Kepler\ FOV by identifying most CVs within a defined magnitude/colour range. The \Kepler\ field is particularly appealing to infer the space density of CVs, since in order to achieve this, orbital periods (required to infer distances, see \citealt{knigge_donor}) and CV sub-type information are required. \Kepler\ provides a unique way to obtain this information for all CVs in the field simultaneously. However, in order to obtain reliable space density estimates for CVs, the CV completeness rate must be both well-known and high within the survey. To this end, it is not yet possible with the follow-up observations presented here to obtain a reliable completeness rate, mainly due to the low number of CVs previously known in the field. We can however take as a reference the 10 known CVs published in \cite{still10}. Out of these, 7 fall within the KIS observations, but only 2 lie within the searched magnitude range of $16<m_{U}<19$. These are V1504 Cyg which is part of our recovered sample of CVs and V523 Lyr which did not pass the $(r-H\alpha)$ cut (but passed the $U-g$ cut). In future, a more lenient $(r-H\alpha)$ cut will have to be employed in order to also recover this CV and others with similar spectra.

\section{Conclusion}\label{sec:conclusion}

In this paper the initial results from an ongoing spectroscopic program on the WHT to identify CVs within the \Kepler\ FOV have been presented. The photometric candidate selection method uses the KIS catalogue (\citealt{greiss}) and selects intrinsically blue objects also displaying $H\alpha$ excess. Of the $44$ candidates in the photometric sample with $m_U<19$, two are already known CVs, whilst 11 are presented here for the first time, resulting in a CV recovery rate of $29\%$. This raises the number of known CVs in the \Kepler\ field by nearly $50\%$ to $\approx25$. Additionally, 13 new QSOs have been identified in the field, of which one displays strong radio emission extended on arcsecond scales. Furthermore 7 F-G-K stars have been presented. An improvement on the candidate selection method to find CVs has been shown by including the WISE colours to the KIS colours ($U$, $g$, $r$, $i$ and $H\alpha$), which greatly enhance the recovery rate for photometrically selected CV and QSO candidates. Whilst 7 objects remain unidentified from the obtained featureless spectra, their WISE colours suggest that these are also QSOs, yielding a QSO recovery rate $\ge29\%$. The identification of new CVs in the \Kepler\ field presented here will allow for in-depth studies of the timing properties for individual systems (see also \citealt{still10,cannizzo10,wood11,scaringi_mvlyra,scaringi_mvlyrb}).

\section*{Acknowledgements}
This research has made use of NASA's Astrophysics Data System Bibliographic Services. S.S. acknowledges funding from NWO project 600.065.140.08N306 to P.J. Groot. S.S. also acknowledges useful and insightful discussions with T. Maccarone and H. Falcke as well as the support astronomer on the WHT J. McCormac. The authors acknowledge the usefull comments provided by the referee which have improved this manuscript.

\bibliographystyle{mn}
\bibliography{KeplerID}

\label{lastpage}

\end{document}